\documentclass[11pt,preprint]{aastex}
\begin{document}

\title{Interaction of a Moreton/EIT wave and a coronal hole}

\author{Astrid~M. Veronig}
\affil{IGAM/Institute of Physics, University of Graz,
Universit\"atsplatz 5, A-8010 Graz, Austria; asv@igam.uni-graz.at}

\author{Manuela~Temmer and Bojan~Vr\v{s}nak}
\affil{Hvar Observatory, Faculty of Geodesy, Ka\v ci\' ceva 26,
HR-10000 Zagreb, Croatia; manuela.temmer@uni-graz.at, bvrsnak@geodet.geof.hr}

\and

\author{Julia K. Thalmann}
\affil{IGAM/Institute of Physics, University of Graz,
Universit\"atsplatz 5, A-8010 Graz, Austria; jut@igam.uni-graz.at}

\begin{abstract}
We report high-cadence H$\alpha$ observations of a distinct Moreton wave observed at
Kanzelh\"ohe Solar Observatory associated with the 3B/X3.8 flare and CME event of 2005 January 17.
The Moreton wave can be identified in about 40~H$\alpha$ frames over a period of 7~min. The
EIT wave is observed in only one frame but the derived propagation distance is
close to that of the simultaneously measured Moreton wave fronts indicating that they
are closely associated phenomena. The large angular extent of the Moreton wave 
allows us to study the wave kinematics in different
propagation directions with respect to the location of a polar coronal hole (CH). In particular we
find that the wave segment whose propagation direction is perpendicular to the CH boundary
(``frontal encounter'') is stopped by the CH which is in accordance with
observations reported from EIT waves \citep{thompson98}. However, we also find that at a
tongue-shaped edge of the coronal hole, where the front orientation is perpendicular to
the CH boundary (the wave ``slides along'' the boundary), the wave signatures can be found up to
100~Mm inside the CH. These findings are briefly discussed in the frame of recent
modeling results.
\end{abstract}

\section{Introduction}

Moreton waves are arc-like propagating disturbances observed in
chromospheric H$\alpha$ images in association with solar flare/CME events
\citep{moreton60,athay61}. \cite{uchida68} developed the theory that the Moreton waves
are just the surface track of a fast-mode MHD wave which is coronal in nature.
When the coronal wave front sweeps over the chromosphere, the enhanced pressure behind the wave front 
compresses the plasma which is observed as the typical down-up swing in H$\alpha$
filtergrams. In recent years, the Extreme-ultraviolet Imaging Telescope (EIT) onboard
the Solar and Heliospheric Observatory (SoHO) directly imaged many globally
propagating disturbances in the low corona \citep[e.g.][]{thompson98,thompson99}, so-called
EIT waves. Soon thereafter, coronal waves were also imaged in soft X-rays
by the SXT/Yohkoh \citep[e.g.][]{khan02, hudson03} and SXI/GOES-12  \citep{warmuth05b, vrsnak06}
instruments. Whether the EIT (and soft X-ray) waves are the coronal counterpart of Moreton waves as
anticipated by Uchida's model is still a matter of debate \citep[e.g.][]{vrsnak05,chen05},
in particular as EIT waves are on average a factor 2--3 slower than Moreton waves
(\citeauthor{klassen00}\ \citeyear{klassen00}; for a discussion on statistical selection effects
see \citeauthor{warmuth01}\ \citeyear{warmuth01}). However, in cases where both the chromospheric
Moreton and the coronal EIT wave were observed, they were lying on closely related
kinematical curves providing strong evidence that, at least in these events,
the Moreton waves are indeed the chromospheric counterpart of the coronal EIT waves
\citep{thompson99,warmuth01,biesecker02,vrsnak02,gilbert04,warmuth04,vrsnak06}.

Some observations show that EIT waves tend to avoid active regions
\cite[ARs;][]{thompson99,wills99} and stop at the boundaries of coronal holes \cite[CHs;][]{thompson98}
as well as near the separatrix between ARs where they may appear as a ``stationary" front
\citep{delannee99}. These observations were reproduced in numerical simulations treating EIT waves
as fast-mode MHD waves \citep{wang00,wu01,ofman02} showing that the wave undergoes strong refraction 
and deflection when interacting with ARs and CHs, which are both high Alfv\'en velocity regions 
compared to the quiet Sun. \cite{delannee99} and \cite{chen02,chen05} proposed different models where
EIT waves are not real waves but the signature of a propagating perturbation related to magnetic
field line opening in the wake of the associated CME.

Observational studies of the interaction of coronal waves with ARs and CHs
are rare and usually hampered by the low cadence of the EIT instrument ($\sim$12--15~min), 
which moreover restricts the studies to slow waves.
In this paper we report high cadence H$\alpha$ observations of a fast Moreton wave
with well defined fronts extending over a large angular sector of $\sim$130$^{\circ}$
which can be followed up to $\sim$500~Mm from the wave ignition center, i.e.\ considerably farther
out than Moreton waves are usually observed. Due to the 12-min cadence, the
EIT wave can be observed solely in one single frame but is roughly co-spatial with
the simultaneously observed Moreton wave fronts as expected from Uchida's theory.
The high-cadence H$\alpha$ observations of this distinct Moreton wave provide us with the possibility
to study in detail the interaction of the coronal wave as traced by its chromospheric ground track 
and a polar CH, which cannot be directly accessed in such detail with present coronal imagers.

\section{Data and Observations}

The Moreton wave associated with the 3B/X3.8 flare of 2005 January 17
(N15,W25; peak time $\sim$09:50~UT) was observed with high time cadence in full-disk H$\alpha$
filtergrams at the Kanzelh\"ohe Solar Observatory \cite[KSO;][]{OtrubaPoetzi03}.
KSO routinely takes full-disk H$\alpha$ images with a
cadence of $\sim$5--10~s and a spatial resolution of 2\farcs2/pixel.
When the flare-mode is triggered, additionally images in the red and blue wing
of the H$\alpha$ spectral line (at the off-band center wavelengths
of H$\alpha + 0.4$~{\AA} and H$\alpha - 0.3$~{\AA}, respectively)
are taken with a cadence of $\sim$60--80~s in each wing.

\begin{figure}
\epsscale{0.9}
\plotone{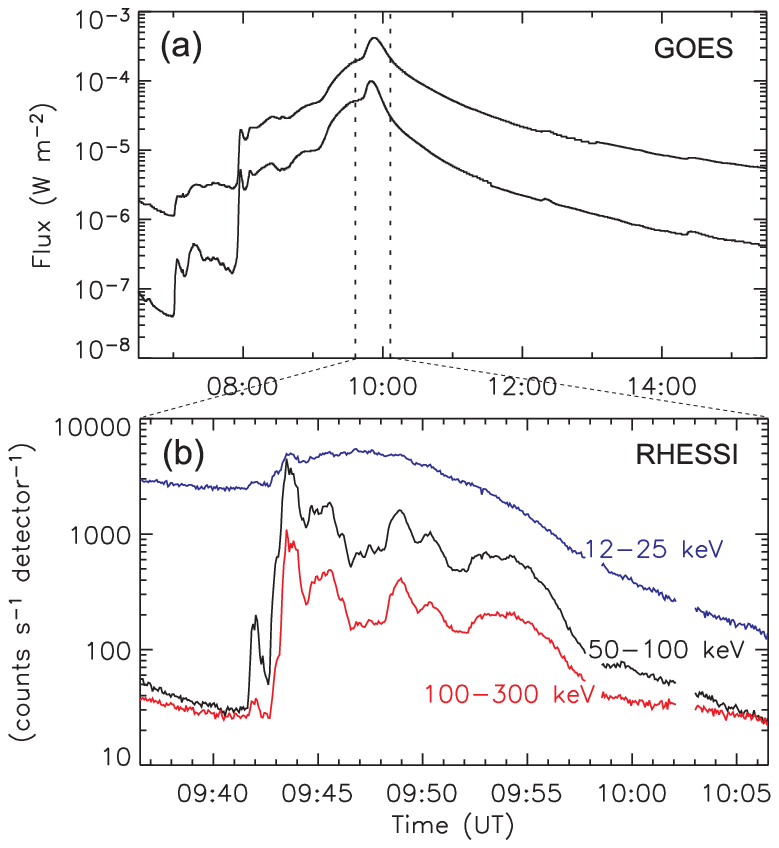} \caption{a): GOES 0.5--4 and 1--8~{\AA} flux.
b)~RHESSI light curves in the 12--25, 50--100, and 100--300~keV
energy bands (4-s integration).
[See the electronic edition of the Journal for a color version of this figure.]}
\label{goes_rhessi}
\end{figure}

The flare evolution is complex and consists of several stages.
The GOES flux (Fig.~\ref{goes_rhessi}a) shows a sudden increase to M2 level
around 8:00~UT, then gradually (over $\sim$100~min) increases further
to the X2 level, and finally shows an impulsive enhancement at 09:40~UT which reaches
the X4 peak at $\sim$09:50~UT. The Moreton wave is associated with this
last peak. For this phase, also hard X-ray data from the Reuven Ramaty High Energy Solar
Spectroscopic Imager (RHESSI; Lin et al.\ \citeyear{LinEA02}) are available. Between
about 09:41 and 09:58~UT spiky emission at high X-ray energies is observed
(Figure~\ref{goes_rhessi}b).

% Fig1

\begin{figure}
\epsscale{0.99}
\plotone{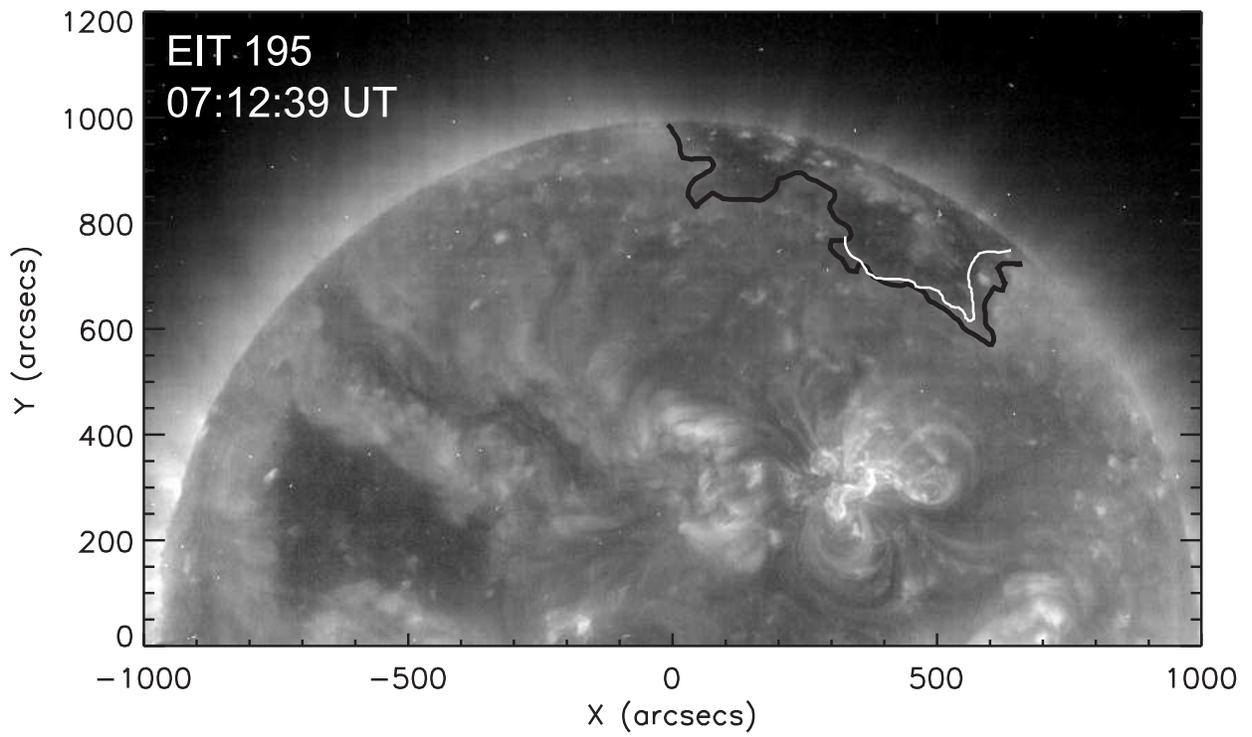}
\caption{EIT 195~{\AA} image of the northern hemisphere of the Sun.
The polar CH boundaries are outlined. The black line indicates the outer, the white
line the inner estimate of the CH boundary.}
\label{ch-det}
\end{figure}

We also checked images from the Extreme-ultraviolet Imaging Telescope
(EIT; Delaboudini{\`e}re et al.\ \citeyear{delab95}) for wave signatures
and the location of the CH boundaries. Figure~\ref{ch-det}
shows an EIT Fe\,{\sc xii}~195\,{\AA} (peak formation temperature of 1.6~MK) preflare
image of the northern hemisphere of the Sun. The boundary of the polar CH,
which is in the direction of propagation of the Moreton/EIT wave, is outlined on the
image. Since the exact CH boundary is difficult to determine, we draw both the outer edge
as well as an estimate of the diffuse inner edge of the CH boundary.

% Fig2

\section{Results}

\begin{figure}
\epsscale{0.7}
 \plotone{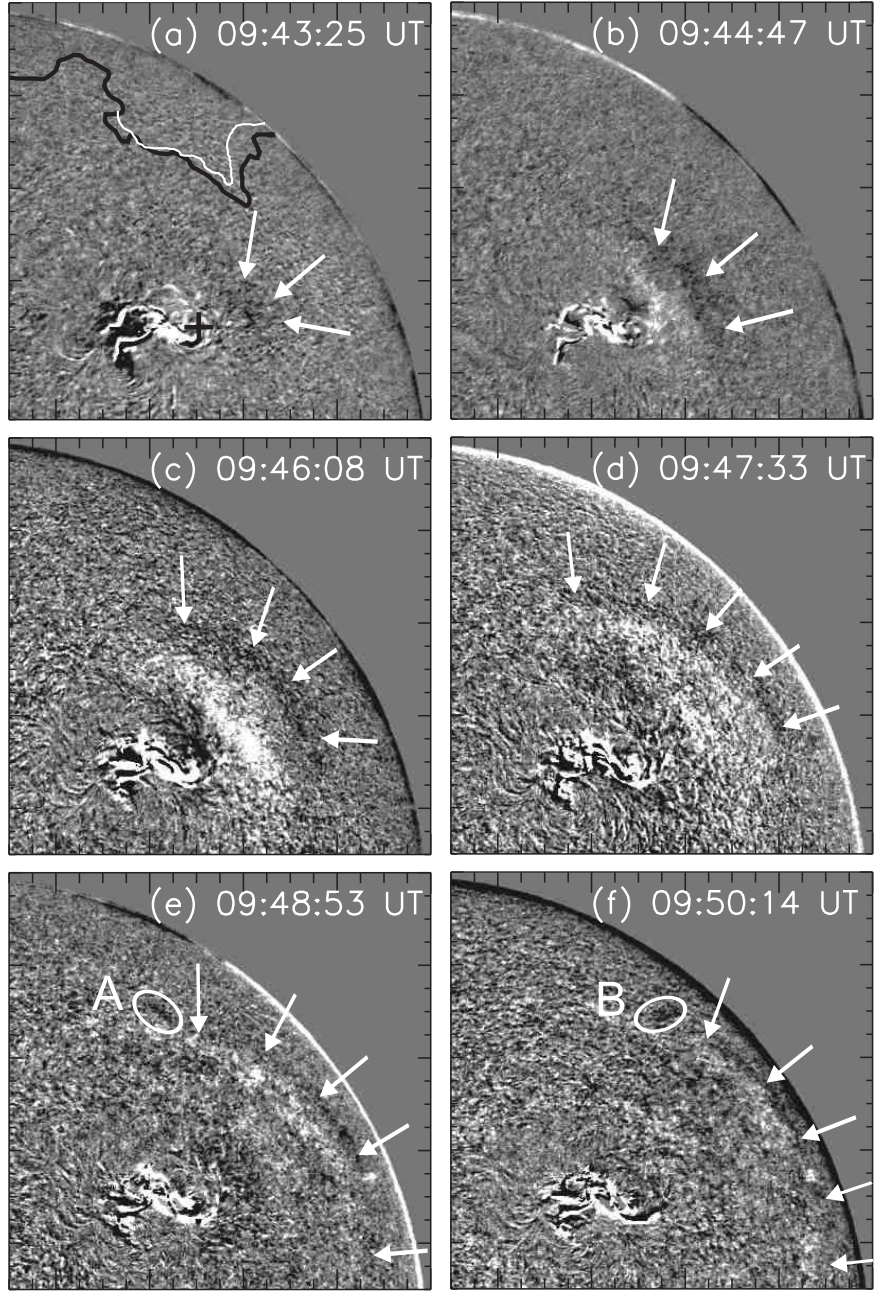}
  \caption{Sequence of H$\alpha+0.4$~{\AA} running difference images. On both axes, the
  plotted field of view extends from $100''$ to $1000''$ from Sun center. The leading edges of the
  Moreton wave fronts (seen as dark fronts in H$\alpha$ red wing observations) are indicated by arrows.
  In panel~a, the determined wave ignition center is indicated by a cross; the CH boundaries are marked
  by white and black lines for the inner and outer estimates, respectively. In panels~e and f, 
  activated features are indicated as ``A" and ``B", respectively.
  [See the electronic edition of the Journal for a color version of this figure.]
  }
    \label{wave_fronts}
\end{figure}

\begin{figure}
\epsscale{0.95}
 \plotone{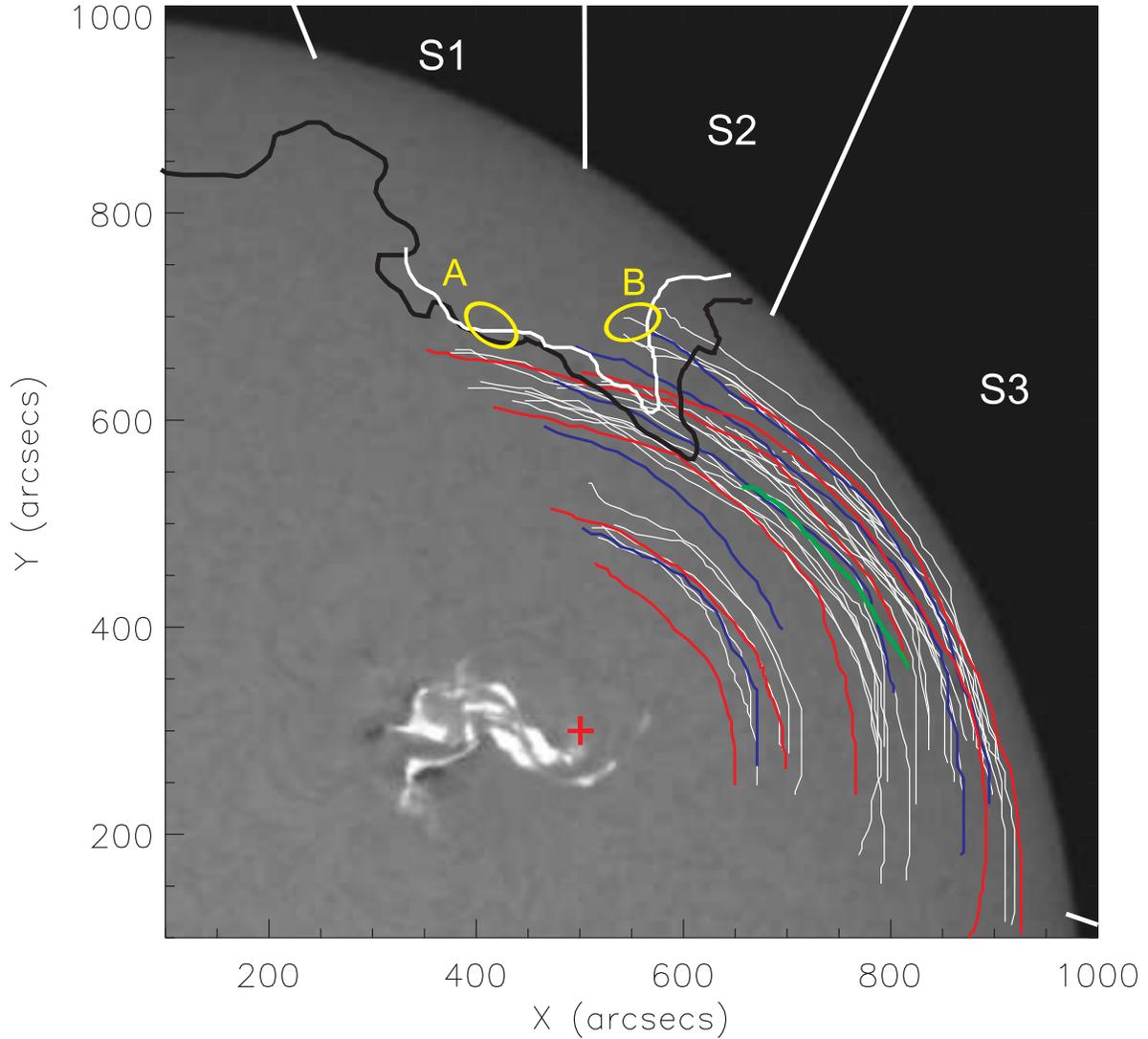}
  \caption{Moreton wave fronts visually determined from H$\alpha$
  (white lines), H$\alpha+0.4$~{\AA} (red lines) and H$\alpha-0.3$~{\AA} (blue lines)
  filtergrams plotted on an H$\alpha+0.4$~{\AA} flare image acquired at 09:47:33~UT.
  The EIT wave front is plotted in green color.
  The polar CH boundaries derived from the EIT image in Fig.~\ref{ch-det}
  are also indicated. The cross marks the determined ignition center of the wave.
  The yellow ellipses indicate the activated features ``A" and ``B".
  The three propagation sectors considered in the kinematical study are indicated
  by straight lines radial to the wave ignition center;
  the sectors are denoted as ``S1", ``S2" and ``S3".
   }
    \label{wave}
\end{figure}

In total, the Moreton wave fronts could be identified in $\sim$30 images taken in the
center, 6~images acquired in the red and 5~images in the blue wing of the H$\alpha$ spectral
line during the period 09:43:25 to 09:50:38~UT. Figure~\ref{wave_fronts} shows a sequence of
H$\alpha+0.4$~{\AA} running difference images of the Moreton wave. In Figure~\ref{wave},
the leading edges of all visually determined wave fronts are drawn on an H$\alpha$ flare image.
The boundaries of the polar CH (inner and outer estimates) determined from
the EIT~195~{\AA} image shown in Fig.~\ref{ch-det} are
also indicated in Figs.~\ref{wave_fronts}~and~\ref{wave}.

% Fig 3

% Fig4

The Moreton wave fronts extend across a maximum sector of $\sim$130$^{\circ}$.
When approaching the CH, there is a distinct change in the wave propagation characteristics
leading to a distortion of the circular shape of the wavefront which can be directly seen,
e.g., in Figs.~\ref{wave_fronts}d,e. From Figure~\ref{wave}
it is evident that the Moreton wave fronts partially intrude into the CH.
The penetration depth depends on the estimated boundaries of the CH. However,
even for the inner estimate, some wave fronts are observed inside the CH.
Further evidence that the wave was able to intrude into the CH is provided by
a feature located inside the CH (marked as ``B" in Fig.~\ref{wave}) that was
activated at the time when the wave passed its position. The wave also activated another feature (``A")
that is located along the border of the CH.

\begin{figure}
\epsscale{0.99}
 \plotone{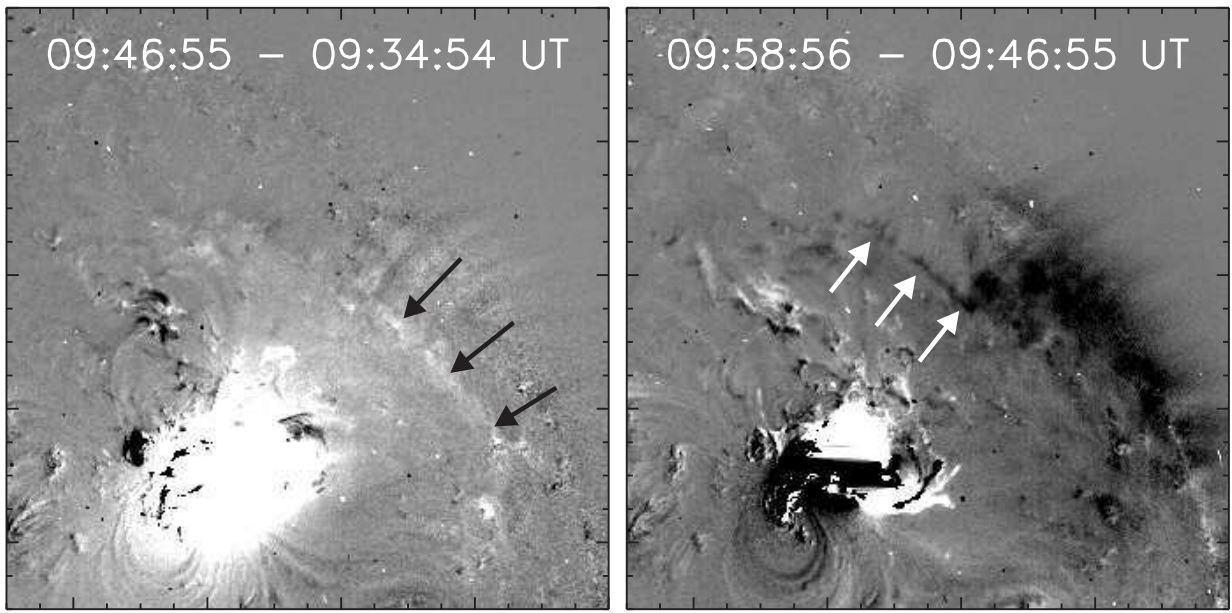}
  \caption{EIT 195~{\AA} running difference images. The left panel shows the EIT wave front
  (indicated by black arrows). The right panel reveals the disappearance of a bright feature
  along the CH boundary (indicated by white arrows) as well as a coronal dimming in the region passed by the wave.
  On both axes, the plotted field of view extends from $100''$ to $1000''$ from Sun center.
   }
    \label{eit_wave}
\end{figure}

Figure~\ref{eit_wave} shows two EIT running difference images. Due to the low cadence of
EIT ($\sim$12~min), there is only one EIT wavefront visible on-disk at 09:46:55~UT (left panel).
In the next EIT frame, the wave has already passed the disk and caused a coronal
dimming behind it (right panel). The figure also reveals that a bright elongated feature
located along the CH boundary (outer estimate) disappeared after the wave passage.
Since this feature reappeared in
later images ($\gtrsim$12~UT), it is possible that it was compressionally heated when
hit by the wave, and was thus no longer visible in the EIT~195~{\AA} passband.
From the EIT observations we find no indication that the wave intruded into the CH.
This may be related to the low image cadence of EIT which does not allow us to get insight into
the EIT wave kinematics, or it may be a real difference between the EIT and H$\alpha$ Moreton wave
propagation characteristics.

\begin{figure}
\epsscale{0.8}
\plotone{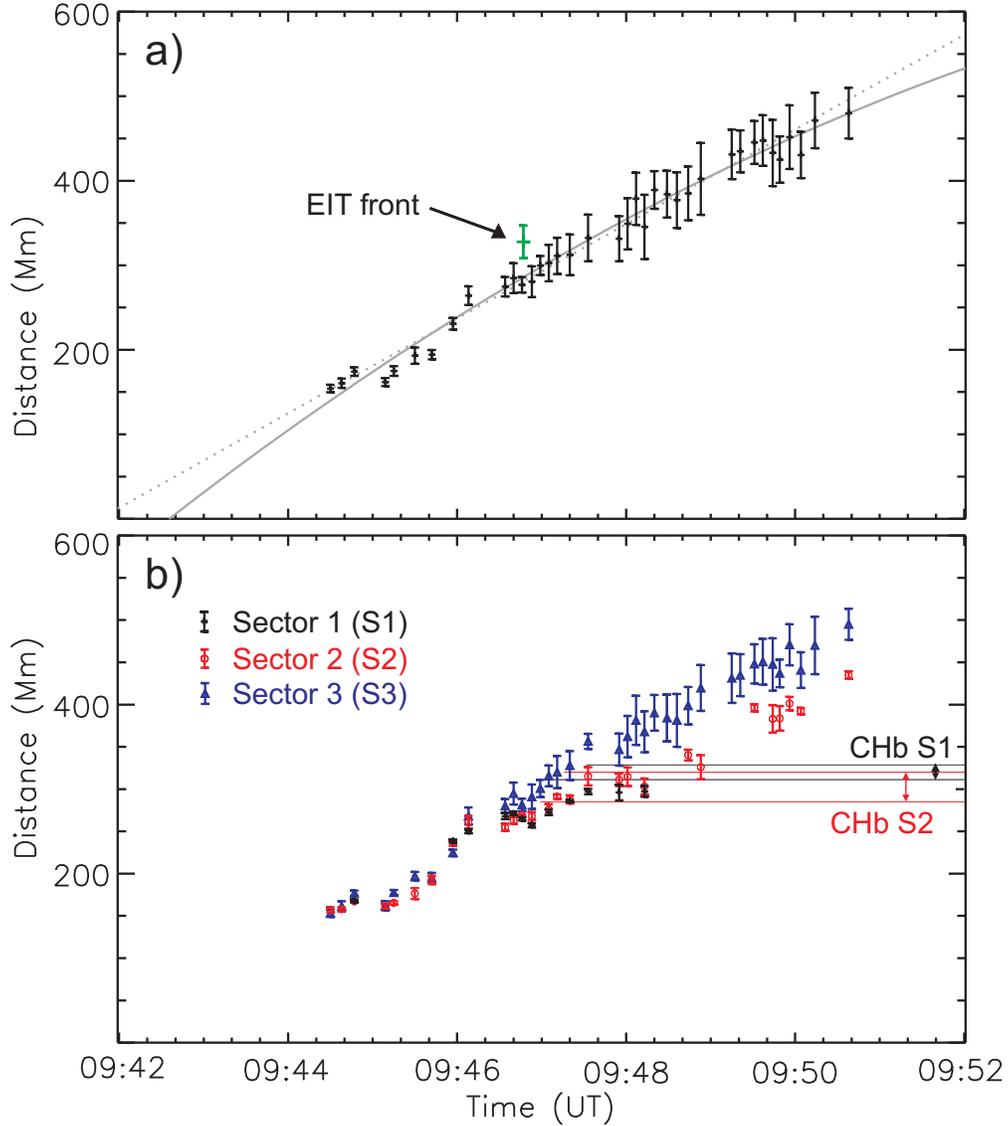} 
\caption{Kinematics of the Moreton wave. a)~Distance vs.\ time
diagram derived from all points on the measured H$\alpha$ wave fronts as well as from the EIT wave front.
The dotted and solid lines indicate the linear and quadratic fit to the H$\alpha$ data, respectively.
b) Distance vs.\ time diagram derived separately for three different propagation
sectors of the Moreton wave: in the direction of the coronal hole (S1, black), in the
direction of the CH tongue (S2, red) and in the direction undisturbed by the CH (S3, blue).
The distance of the CH boundaries into sector S1 (``CHb S1") and towards the tongue
in sector~S2 (``CHb S2") are indicated by the horizontal lines (lower lines -- outer estimates,
upper lines -- inner estimates).
}
\label{kinematics}
\end{figure}

The Moreton wave ignition site was estimated by applying circular fits to the
earliest observed H$\alpha$ wave fronts, whereby the projection effect
due to the spherical solar surface was taken into account (see Warmuth et al.\
\citeyear{warmuth04}). We found the ignition site at $x\approx 505''$ and $y\approx 300''$
from Sun center, which is approximately $50''$ NW from the centroid of the
flare area impulsively activated during this time and outside the strong sunspots of the active
region (cf.\ Fig~\ref{wave}).
We then determined for each point of the wave front, i.e.\ considering its total angular extent,
its distance from the wave ignition site measured along great circles
on the solar surface. The derived time-distance diagram
is plotted in Fig.~\ref{kinematics}a. At each instant the mean value derived
from the distances of all points on one wave front together with the
standard deviation (error bars) is plotted.
The first wave front observed in H$\alpha + 0.4$~{\AA} at 09:43:25~UT
was ignored in the kinematical curves since it was very diffuse and therefore difficult to
measure. In H$\alpha$, the wave propagation can be
observed up to a distance of $\sim$500~Mm from the ignition site.
This is considerably farther out than Moreton waves are usually observed.
Warmuth et al.~(\citeyear{warmuth04}) report an average value for the maximum
distance of 300~Mm.

%Fig5

Also plotted in Fig.~\ref{kinematics}a are the results from a linear and quadratic least squares fit
to the data. From the linear fit we obtain a mean wave speed of $930\pm 30$~km~s$^{-1}$. The quadratic
fit gives a significant deceleration of $-1240\pm 470$~m~s$^{-2}$. From the quadratic fit we
find a back-extrapolated start time of the wave of 09:42:40~UT $\pm~30$~s, from the linear fit it is 
09:41:50~UT $\pm~20$~s. These start times point to the impulsive onset of the flare hard X-ray emission 
observed by RHESSI (see the first peak at 50--100~keV in Fig.~\ref{goes_rhessi}).
The distance derived from the EIT wave front at 09:46:55~UT indicates that the EIT wave is
$\sim$50~Mm ahead but roughly co-spatial with the instantaneously observed Moreton wave fronts.
This finding is in contrast to the statistical result that EIT waves are significantly (by a factor 2--3)
slower than Moreton waves \citep{klassen00} but agrees with Uchida's sweeping skirt hypothesis
in which the H$\alpha$ Moreton wave is the chromospheric surface track of the coronal wave as observed
by the EIT, SXT and SXI instruments.

In Fig.~\ref{kinematics}b, we present the time-distance diagram separately for three propagation
sectors. Sector~1 includes the wave segment where the wave front and CH boundary are quasi parallel
(``frontal encounter'' of the wave). It is defined by the angular sector $[90^\circ,110^\circ]$ measured from the
$x$-axis with the wave center as origin. Sector~2 covers the direction of the western
boundary of the CH (revealing a ``tongue"-like shape) where the wave fronts impinge quasi-parallel
as well as normal to the CH boundaries (defined by $[65^\circ,90^\circ]$).
Sector~3 is in the direction that is unaffected by any CH boundary (defined by $[-20^\circ,65^\circ]$).
The three sectors are indicated in Fig.~\ref{wave}.

Up to about 09:46~UT all three kinematical curves are very similar. Thereafter, the wave segments
of sectors 1 and 2 (i.e.\ in direction of the CH) slow down with respect to the undisturbed direction
(sector~3). At 09:48:20~UT the wave segments in sector~1 stop at a distance of about 300~Mm, i.e.\
closely ahead of the CH boundary (see also Fig.~\ref{wave}). The wave propagating into the direction of the CH tongue
(sector~2) also shows deceleration when approaching the CH, but some of the wavefronts
intrude into the CH. At that time their propagation characteristics (inside the CH) are different
from that of the wave segments propagating into sector~3, i.e.\ the undisturbed direction.
In particular, their distance to the wave ignition center is about 60--90~Mm smaller than that of
the wave front segments not disturbed by the CH, indicating that the wave fronts are significantly distorted
by the presence of the CH.

%Fig6

\section{Discussion and Conclusions}

High cadence H$\alpha$ observations of a distinct Moreton wave extending over an angular
extent of $\sim$130$^{\circ}$ allowed us to study the propagation characteristics in
different directions with regard to the presence of a polar CH.
In particular we find that in the frontal encounter (sector~1), the wave stops at the CH boundary.
On the other hand, at the western side of the CH where the front orientation is normal to the CH boundary,
the wave can to a certain degree intrude into the CH. Depending on the different (i.e.\ inner/outer) CH
boundary estimates we find a maximum penetration depth of 65 and 100~Mm, respectively.
From the EIT observations we find no indication that the wave was able to intrude into the CH.
This may be related to the low image cadence of EIT --- while the Moreton wave can be measured
in 40 frames, the EIT front is observed only in a single frame --- or it may be a real difference
between the EIT and H$\alpha$ wave propagation characteristics. 

In the MHD fast-mode hypothesis, the velocity of the coronal wave which propagates basically
perpendicular to the magnetic field lines, is determined
by the fast magnetosonic speed $v_{\rm ms} = (v_A^2 + c_s^2)^{1/2}$ with
$v_A = B/(4 \pi \rho)^{1/2}$ the Alfv\'en speed and $c_s = (\gamma k T / \overline{\mu} m_u)^{1/2}$
the sonic speed, where $B$ denotes the magnetic field strength, $\rho$ the mass density, $n$ the particle density,
$T$ the temperature, $\gamma$ the adiabatic exponent, $k$ the Boltzmann constant, $m_u$ the atomic mass unit,
and $\overline{\mu}$ the mean molecular mass. The Moreton wave of 17 January 2005 has its starting point
at the NW border of AR~10720 and propagates through quiet Sun regions, partly into the direction of the
northern polar CH. The higher magnetosonic speed in CHs compared to quiet Sun regions, which causes refraction and reflection of the wave at the CH, is mainly due to the higher Alfv\'en velocity in CHs. The density in CHs is about a
factor~3--10 smaller compared to the quiet Sun \citep{gabriel92,young99,cranmer00}. Using the solar minimum
magnetic field model by \cite{bana98} which is a reasonable representation
of the large scale magnetic field outside ARs, we find that the global field is about a factor~3 smaller
at the latitude of the wave initiation ($\sim$$15^\circ$) than at $\sim$35--40$^\circ$, where the southern
boundary of the CH is located. The temperature in CHs is up to a factor 2 smaller than the ambient coronal plasma.
Thus, the Alfv\'en velocity in CHs is on average a factor 5--9 higher than in the ambient quiet Sun coronal plasma,
whereas the sonic speed is about a factor 1.4 smaller. Assuming that in the corona the Alfv\'en velocity is significantly higher than the sonic speed (low-$\beta$ plasma), this implies that the fast magnetosonic speed in the CH
is about 5--9 times higher than in the quiet corona. As argued by some authors \cite[e.g.][]{wang00}, in quiet regions  the plasma-$\beta$ could be of the order of unity and thus the sonic velocity comparable to the Alfv\'en velocity. However, also in this case the fast magnetosonic speed in the CHs is still considerably higher than in quiet regions, 
namely by a factor of 3--6.

The stopping of the wave at the CH boundary in the case of the ``frontal
encounter'' was numerically simulated by \cite{wang00} and \cite{wu01}. As the wave approaches 
a region of high magnetosonic speed it is refracted away from that region and the wave
propagation is halted \citep{wang00,wu01}. Due to the low values of the plasma-to-magnetic pressure ratio in the
CH region, the mass motion is constrained \citep{wu01} and the plasma piles up when the wave encounters
the CH boundary. That could explain the activated (neutral line) feature ``A''. There could be
also a small transmitted component of the waves as seen in the simulations by \cite{ofman02}. However, the
magnitude of the transmitted component must be much smaller than that of the reflected component, and is 
therefore difficult to detect.

A related phenomenon is the leakage of a fast mode wave into the CH region \citep{wu01}. Our
observations apparently reveal such a leakage at the western edge of the CH, where the wave passed
the CH tongue. After passing the tongue we still see wave signatures within the CH region.
Moreover, the wave caused there an activation of feature ``B''. Since that happened in a different
geometry (the wave is here ``sliding along'' the CH boundary), the activation was most likely due
to the interaction of the wave flank that managed to penetrate into the CH from aside, and the
pre-existing magnetic structure therein. Such a process is possibly similar to the
wave/AR interaction modeled by \cite{ofman02}, since ARs are as CHs high Alfv\'en/magnetosonic 
velocity regions. It was shown by \cite{ofman02} that when the fast magnetosonic wave encounters an AR 
it undergoes strong reflection and refraction. On the other hand, transient currents are induced
within the AR, and the resulting Lorentz force causes dynamical distortion of the magnetic field
structure. The resulting magnetic compression of plasma generates flows which could be seen as an
AR activation, similar to what we observed by the activation of feature B inside the CH.

The present analysis shows the importance of high time cadence observations
to interpret properly the phenomenon of the interaction of a coronal/Moreton wave with a CH. Up to now,
conclusions were drawn only from EIT observations which are taken with low time cadence, and thus
might miss important details.

\acknowledgements We gratefully acknowledge support by the Austrian {\em
Fonds zur F\"orderung der wissen\-schaftlichen Forschung} (FWF) under grants
P15344 and J2512-N02. We thank the anonymous referee for the suggested improvements 
to the paper.

\clearpage

\clearpage

\clearpage

\clearpage

\clearpage

\clearpage

\end{document}